\documentclass[3p,twocolumn]{elsarticle}
\usepackage{graphicx,rotating}
\usepackage{lineno}
 
\begin{document}

\title{Lineshape response of plastic scintillator to pair production of
4.44 MeV $\gamma$'s}

\begin{abstract}
  We measure the distribution of energy deposited in a 40x88 mm plastic
  scintillator by $e^+ e^-$ pair production of 4.44 MeV $\gamma$-rays. We observe
  the double-escape peak of 3.42 MeV from pair production by tagging 511 keV
  annihilation radiation in two high-Z scintillators.
The source is a standard commercial
neutron source using $\alpha$-emitting $^{241}$Am encapsulated with $^{9}$Be, which has a
reaction branch feeding the first I$^{\pi}$=$2^+$ state of $^{12}$C making the 4.44 MeV $\gamma$-rays.
We demonstrate the extraction of the double-escape peak from the large neutron-produced backgound, and explore
some of the features and difficulties of this technique with our apparatus.

\end{abstract}

\author[1,2]{Melisa Ozen}

\author[1,3]{John A. Behr
}
\ead{behr@triumf.ca}

\author[3]{Michelle Khoo}

\author[3]{Felix Klose}

\author[1]{Alexandre Gorelov}

\author[4,5]{Dan Melconian}

\affiliation[1]{organization={TRIUMF},
  addressline={4004 Wesbrook Mall},
  city={Vancouver, BC},
  postcode={ V6T 2A3},
  country={Canada}
}

\affiliation[2]{organization={University of Ottawa},
  addressline={Dept. of Physics},
  city={Ottawa, ON},
  postcode={ K1N 6N5},
  country={Canada}
}

\affiliation[3]{organization={University of British Columbia},
  addressline={Dept. of Physics and Astronomy},
   city={Vancouver, BC},
  postcode={ V6T 1Z1},
  country={Canada}
}

\affiliation[4]{organization={Texas A\& M},
  addressline={Cyclotron Institute},
  city={College Station, Texas},
  postcode={ 77843-3366},
  country={U.S.A.}
}

\affiliation[5]{organization={Texas A\& M},
  addressline={Department of Physics and Astronomy},
  city={College Station, Texas},
  postcode={ 77843-4242},
  country={U.S.A.}}







\maketitle


\section{Introduction}

Large plastic scintillators can be used for $\beta$ energy spectroscopy~\cite{Fenker2018,Minamisono2011,Clifford1983}, because the relatively low Z minimizes energy loss from $\beta$ backscattering and bremsstrahlung. To understand the experimental $\beta$ spectrum then requires a convolution of a theoretical spectrum with measured or simulated energy response of the detector. An exemplary experiment studied the energy response of plastic scintillator in detail as a function of $\beta$ energy from 0.8-3.8 MeV
using monoenergetic $\beta$'s selected by a magnetic spectrometer~\cite{Clifford1984}. Here we describe features of a simpler technique, and use it to characterize the energy response of plastic scintillator at one relatively high energy. 

Sources using ($\alpha$,n) reactions on $^{9}$Be and $^{13}$C are neutron
calibration standards, and both neutron spectra~\cite{Lorch1973} and
high-energy $\gamma$ production~\cite{Scherzinger2017} have been well-characterized in the literature. These sources are routinely used to calibrate large high-Z scintillators at $\gamma$ energies 4.44 and 6.13 MeV produced by population of
excited states of $^{12}$C and $^{16}$O, as the large pair production cross-sections and photopeak fractions can overcome events from inelastic scattering of fast neutrons and from thermal absorption, even in singles. We do not find in the literature use of
these $\gamma$'s to calibrate plastic scintillator. 

In our relatively large
plastic scintillator, approximately 4\% of 4.44 MeV $\gamma$'s will Compton scatter, while approximately 0.5\% will pair produce.
We find here that timing and energy tags for both 511 keV $\gamma$'s in back-to-back high-Z GAGG scintillators
adequately selects the double-escape 3.42 MeV energy peak from the stopped pair of electron and positron, and we address problems from remaining backgrounds from neutron-produced events.

We use this technique to measure three properties needed for $\beta$ spectroscopy:
the centroid energy of the double-escape peak, to test linearity of the energy
response;
the peak's full width at half maximum (FWHM), to test whether the energy resolution is determined by photon statistics together with dark current;
the number of events with substantial energy missing, which in plastic
scintillator is typically determined by bremsstrahlung of the $e^+ e^-$ pair.

This latter lineshape tail is critical for any $\beta$ experiment that does not use a magnetic
spectrometer, as it mischaracterizes higher-energy $\beta$ strength as lower energy.
Modern Monte Carlo techniques can usually simulate that lineshape tail to within about 10\% of its value, setting a goal for our measurements of the lineshape tail/peak ratio.
Unfortunately, our observed lineshape response tail greatly exceeds a Monte Carlo
simulation of electromagnetic processes. 
We can qualitatively explain the tail from observed backgrounds likely
involving neutrons from the source, a serious limitation of our implementation of this technique.

\section{Experiment}

\begin{figure}[htb]
 \includegraphics[width=\linewidth]{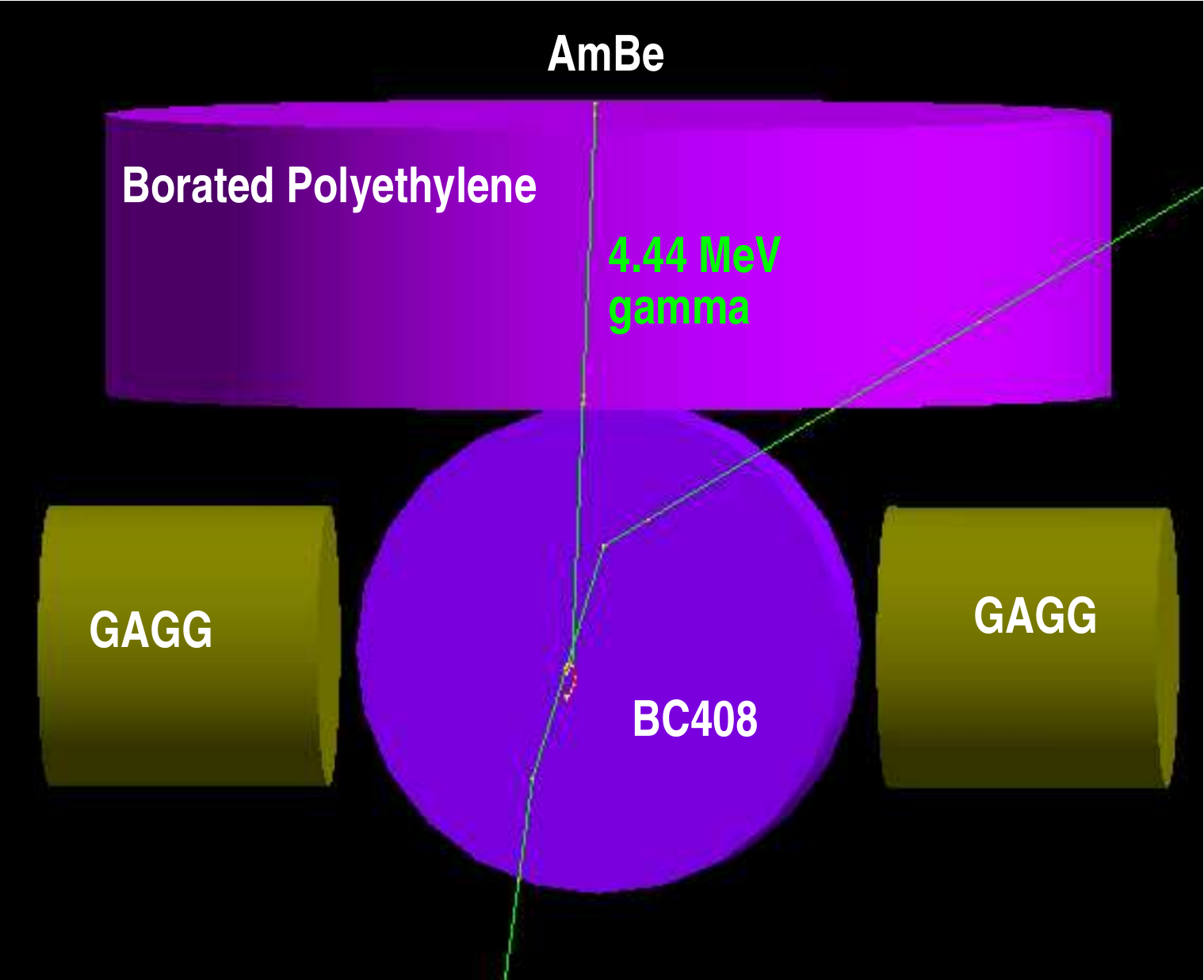}
  \caption{
    GEANT4 scale model of the experimental geometry, showing 40x88 mm plastic scintillator, two 5x5cm GAGG scintillators, AmBe source location, and borated polyethylene shielding. A pair production event is shown where both annihilation $\gamma$'s miss the GAGG's.
  }
  \label{fig-geometry}
\end{figure}

Our specific motivation is that we have recently
changed scintillator readout from photomultiplier tube (PMT) to silicon photomultiplier (SiPM) technology, to minimize gain changes from the switching magnetic fields
of our magneto-optical trap. This created a  need to test the energy response.

Figure~\ref{fig-geometry} shows the geometry, with back-to-back 5x5 cm high-Z gadolinium aluminum
garnet (GAGG) scintillators to detect the 511 keV $\gamma$'s from pair production. The GAGG
detectors have high photopeak fraction of 80-90\% at 511 keV.
The plastic scintillator is BC408. The sides are wrapped in about 10 layers of 0.003'' thick teflon (i.e. about 0.75 mm).
The front face is covered by nitrocellulose 0.12 mm thick, which
in contrast with teflon~\cite{Ghosh2020,Haefner2017} is known to be highly reflective at this thickness~\cite{nitrocellulose},  minimizing $\beta$ energy loss and straggling.
The photons are collected by a SensL ARRAYC-60035-64P-PCB 57x57mm silicon photomultiplier (SiPM), and read out by commercial electronics.
The back surface area that is
not covered by SiPM is covered in 0.75 mm thick teflon. 

Large-area SiPM readout is used for all detectors. We caution that at first we saw large changes in bias currents of up
to a factor of three in all
three detectors in 12 hours of counting. Our $^{241}$Am+Be (AmBe) source has 1x10$^6$ $\alpha$ decays/s and produces about 5 $\mu$Sv/hr neutron doses at 0.3 m distance.
Extrapolations from published
SiPM effects from neutrons~\cite{Musienko2017,Tsang2016} suggested that would not be an
issue as we estimated 1/100 of fluences that produced damage, yet the
bias currents stopped increasing when we added 5 cm of borated polyethelyne
shielding. At least one prominent $\gamma$ line in the GAGG's above and near 511 keV
was reduced to negligible amounts by the neutron shielding.
We were
somewhat surprised that the very high neutron capture cross-sections of
gadolinium isotopes are not producing overwhelming $\gamma$-ray backgrounds.

We list here the sources we used for low-energy calibration.
The internal conversion
electrons from $^{207}$Bi decay produced instead of the 1064 keV $\gamma$,
averaging 995 keV electron energy,
provide an accurate calibration with considerable precision on the energy resolution at this low energy of interest.
We also use $\gamma$-rays from that 1064 keV $^{207}$Bi $\gamma$,
the 662 keV $\gamma$ from decay of $^{137}$Cs, and
the 1462 keV $\gamma$ from naturally occuring $^{40}$K decay,
producing Compton edges at 858, 478, and 1245 keV.

\subsection{Coincidence data and cuts}

\subsubsection{Relative timing}

\begin{figure}[htb]
  \includegraphics[angle=90,width=\linewidth]{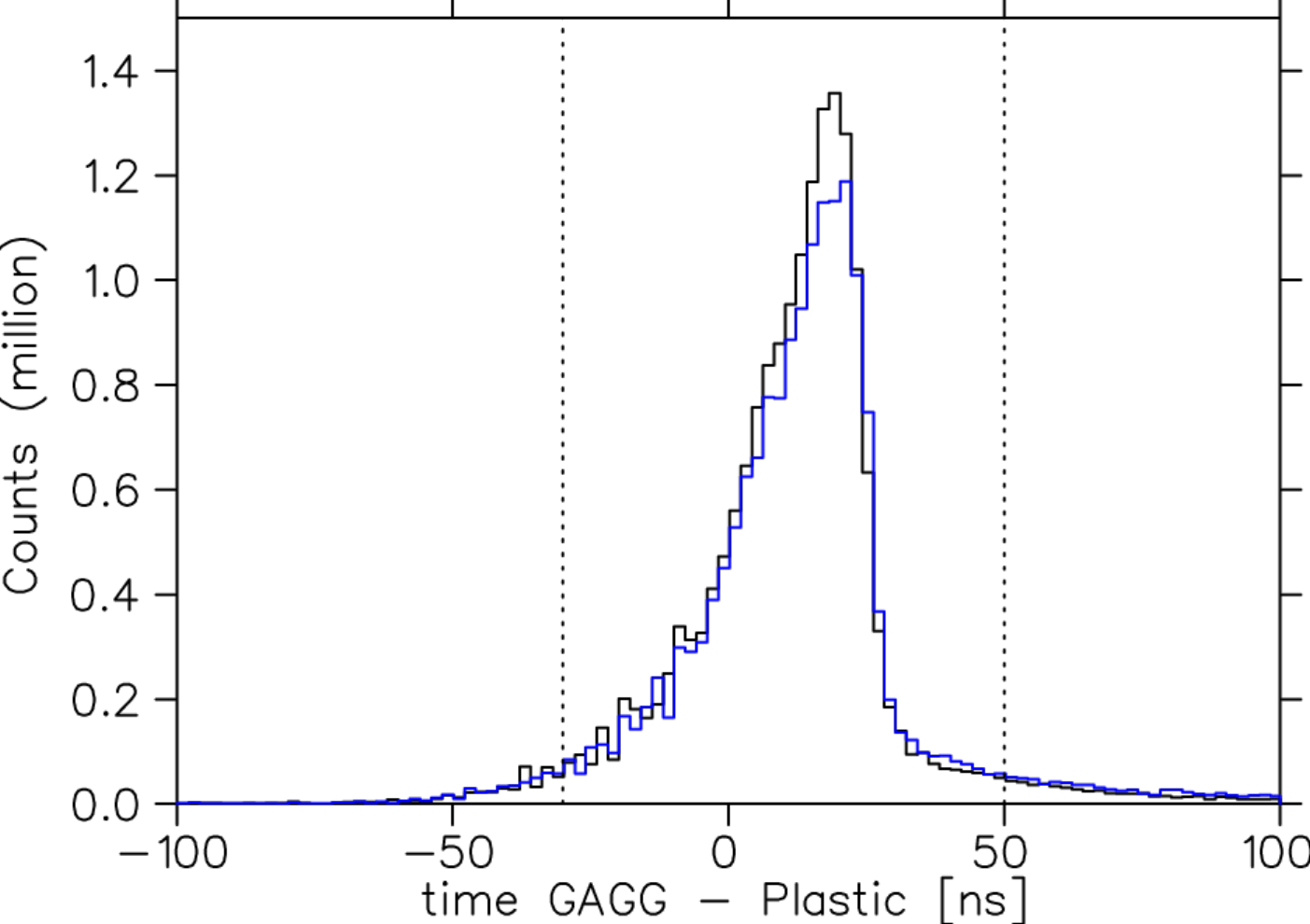}
  \caption{
    Representative timing resolution of GAGG detectors with respect to the plastic scintillator (from 16 hr of data). Timing cut is indicated.
      }
  \label{fig-timing}
\end{figure}

Conventional constant fraction timing produces relative timing spectra
like the one in Figure~\ref{fig-timing},
characterized by FWHM of 20 ns. This is sufficient to remove events from
thermalized neutrons and from accidental coincidences, but leaves 
events in the GAGG energy spectra that are likely from nonthermalized neutrons.
Spectra below are based on  $\pm$ 2$\sigma$ cuts in timing, taking
about 95\% of the total events.
We have studied both more and less restrictive cuts, and find that tail and peak in the plastic response are qualitatively similar.

Neutrons with 5 MeV energy need 2 ns to travel 5 cm, so timing more than an order of magnitude better would be needed to reject fast neutrons from the AmBe source.
The readout of our large-area SiPM's and their relatively large capacitance limits the 10-90\% risetime to  
40$\pm$3 ns in the plastic, and contibutes part of the observed risetime of
86$\pm$3 ns in the GAGG, so timing improvements would need much different readout and likely faster high-Z scintillators.

\subsubsection{$\gamma$ Energy}

\begin{figure}[htb]
  \includegraphics[angle=0,width=\linewidth]{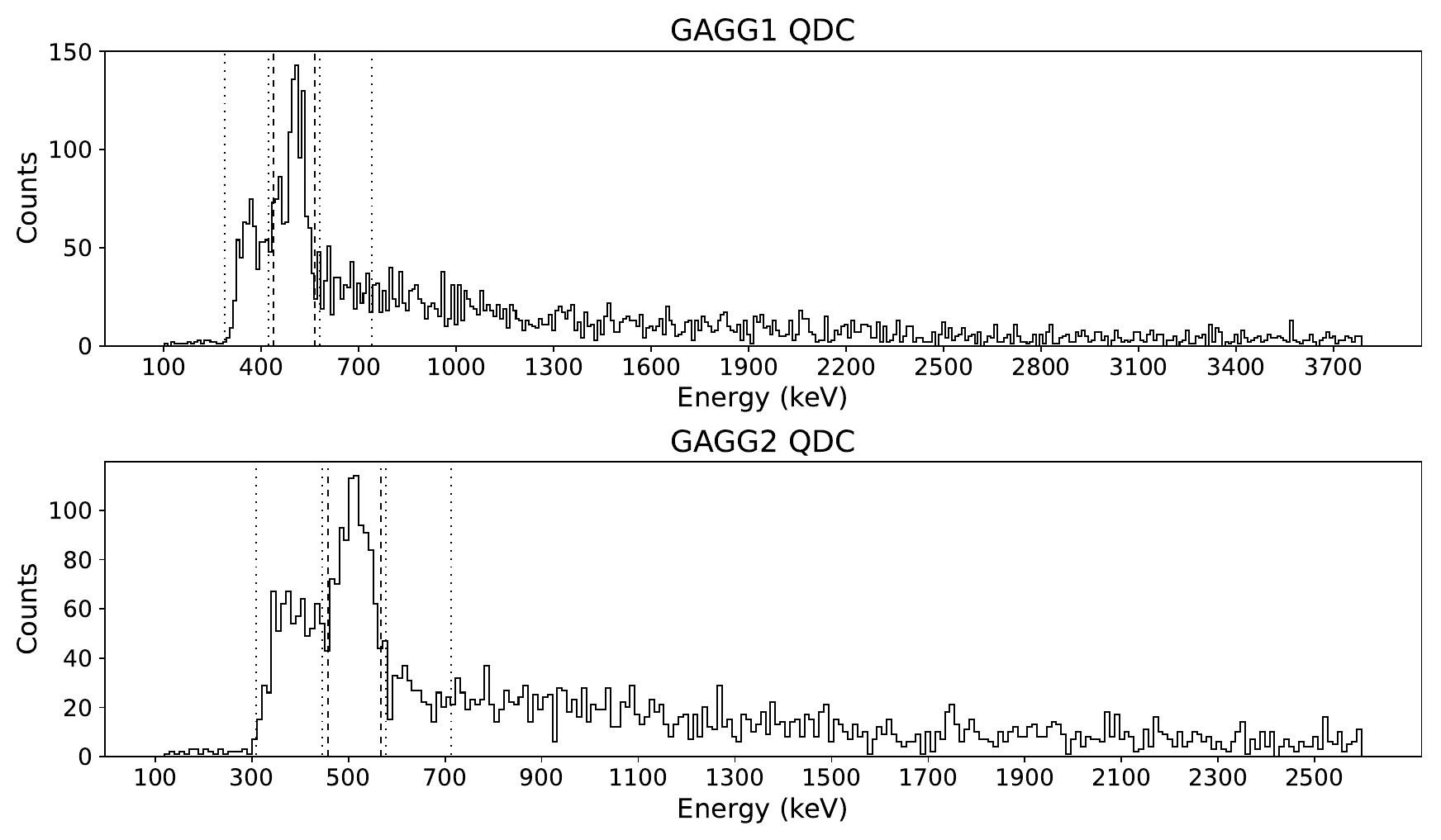}
  \caption{
    Energy spectra of the GAGG detectors in timing coincidence with the
    plastic scintillator (note energy calibration and resolution are different). Cuts of $\pm 2 \sigma$ are shown around the annihilation peaks, along with background regions extending the next 5 $\sigma$ above and below the signal.
          }
  \label{fig-GAGG}
\end{figure}

The GAGG energy spectra in timing coincidence with the plastic
are shown in Figure~\ref{fig-GAGG}. Clear 511 keV annihilation photopeaks
are seen. All 88 hours of data collection is shown.

There are more events at lower energy than the 511 keV photopeaks than
would be expected from Compton edges. Thus by inspection
there are also backgrounds continuous in energy extending
under those photopeaks, both at lower and higher energy.
We attribute these to possible events
associated with neutrons not excluded by our coincidence timing cut.
We will show the plastic scintillator energy spectrum for these background
events in the next section.

\subsection{Energy spectrum of plastic scintillator}

\begin{figure}[htb]


  \includegraphics[angle=90,width=\linewidth]{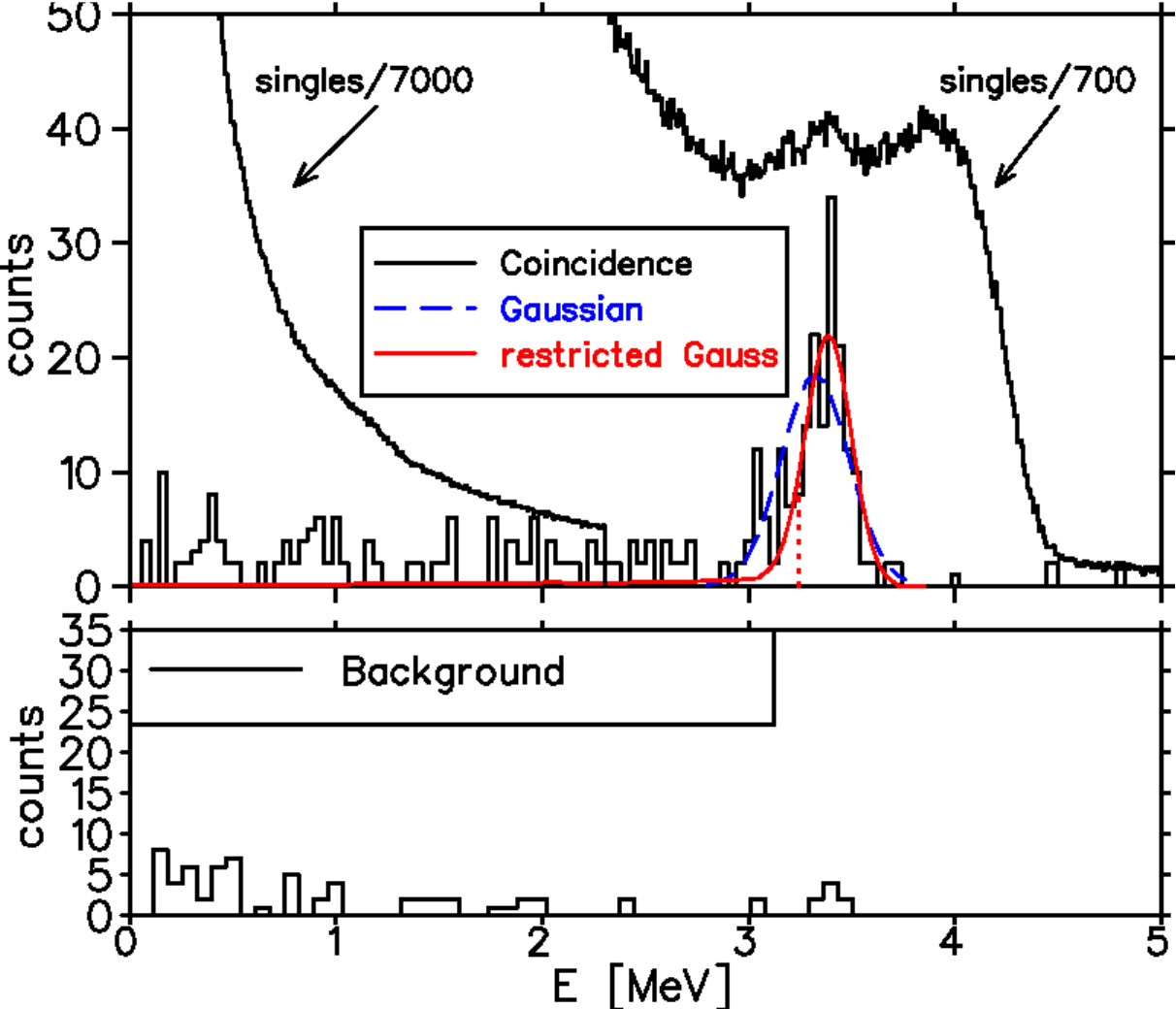}


  
  \caption{
Top:   The main technique demonstration: energy spectrum of all plastic scintillator events
    in timing coincidence with both GAGG's, and requiring annihilation radiation photopeaks in
    both GAGG's. Singles spectrum, maximum likelihood fits, and GEANT4 tail simulation are discussed in text. Energy scale is determined in the Energy Calibration subsubsection.\\
    Bottom: The energy spectrum for background events in the
    plastic scintillator for events meeting timing cuts with 511 keV in one GAGG but
    more or less energy in the other GAGG. 
          }
  \label{fig-plastic}
\end{figure}

Fig.~\ref{fig-plastic} shows the singles plastic scintillator energy spectrum, with no GAGG coincidence cuts.
Although this spectrum is dominated by neutron-produced events, there are peaked structures at the single escape peak energy and the double escape peak energy. 
The singles spectrum, however,  also shows an unanticipated structure about 200 keV lower than the double escape peak, some of which survives the coincidence cuts. We have been unable to identify a physical origin for this extra structure.  It could come from some neutron-produced background. Although data at 995 keV from IC electrons do not show any sign of such an effect, those electrons are confined to within 1 cm of the center front face of the detector. So it has been difficult to rule out some flaw in construction of the scintillator, or light collection from $\gamma$-produced events, since they illuminate the entire detector. 

Fig.~\ref{fig-plastic} also shows the measured plastic scintillator energy with cuts, i.e.
in timing coincidence with GAGG and requiring 511 keV photopeaks in
both GAGG detectors.  The double escape peak at 3.416 MeV is clear. This is the result
of 88 hours of data collection. We consider this a demonstration of our main result, that tagging
on annihilation radiation reveals the double escape peak.

The extra structure seen in singles may survive in coincidence as a tail from 3.0 to 3.2 MeV on this double escape peak, but the statistics in coincidence are poor. 
We note that our
GEANT4 simulation (described below) naturally includes energy loss from e$^+$ or e$^-$ leaking out the edge of the detector, so a small tail with 9\% of the peak events linearly decreases to zero 100 keV away from the full energy peak. Folding this into our Gaussian fits changes neither the centroid nor the FWHM appreciably, and this effect does not explain the extra structure observed.

We show two maximum likelihood fits. A Gaussian fit to the coincidence data above 2.8 MeV has $\sigma$=165$\pm$16 keV, in good agreement with the statistical deviation of the data about its mean, but $\chi^2$/N$_{\rm dof}$ (number of degrees of freedom) is 3.0 with confidence level 4x10$^{-4}$, suggesting one Gaussian is not a good physical model. We do not show our attempt to add an exponential tail, used by Clifford {\em et al.}~\cite{Clifford1984} to model nonuniform light collection, noting only that it does not improve goodness of fit so does not model the extra structure well. Given that the neutron background is steeply falling in singles, and vanishes in coincidence above the double-escape peak, we also see no way to reliably model the neutron events that survive coincidence cuts.

The other maximum likelihood Gaussian fit, restricted to an energy range above the tail, is the best physical description we have found of the energy centroid response of this detector. It has $\sigma$=110$\pm$11 keV, with $\chi^2$/N$_{\rm dof}$=2.4 and confidence level 1\%. The resulting centroid and FWHM are used in subsections below. We show the convolution of this restricted Gaussian fit with our GEANT4 simulation in Fig.~\ref{fig-plastic}.

\subsubsection{Energy calibration}
  
A fit to a linear energy calibration, Fig.~\ref{fig-linearity}, shows a systematic deviation.
If all calibration data were used,
the 3.42 MeV point would show 5\% deviation from linearity with high statistical significance (6$\sigma$), while the fit has large $\chi^2$.
Such a nonlinearity seems unlikely to be due to SiPM response, given their known linearity at these relatively low count rates of a few 100 counts/sec, but could be from other electronics elements.
Note we have included 30 keV minimally ionizing energy loss of the $^{207}$Bi conversion electrons in the 0.12 mm nitrocellulose front face reflector.

A possible physical explanation involves excluding the conversion
electron from $^{207}$Bi from the fit. The resulting fit of the $\gamma$ sources
is linear to better than 0.5\% precision, and the residuals in Fig.~\ref{fig-linearity} show a 3\% deviation
of the 995 keV electron. To show such a deviation, we require considerable precision and accuracy to extract the energy centroid produced by Compton edges, for which we use a differential method from the literature~\cite{differential}.
The fit still has a relatively high $\chi^2$, with probability 0.5\% that a random measurement would show a higher $\chi^2$, which
we attribute to difficulties of Compton edge extraction at precision $\sim$0.1\%. 
%

The higher gain for the conversion electron 
might be explained by
3\% higher light collection
from the first 5 mm of the plastic scintillator where the 995 keV electrons
are stopped, compared to the $\gamma$-ray sources which uniformally
illuminate the entire detector. This is roughly consistent with optical collection simulations of this detector geometry,
which show 5-10\% nonuniformity for electrons deposited in the front face at various radii from center out
towards the cylinder edge.
So different light collection for electrons and $\gamma$'s is intuitively reasonable,
and is a large enough effect to
need a correction to this paper's $\gamma$-produced double escape peak when used to calibrate $\beta$ spectroscopy.
We note also that the known difference of energy loss in scintillators between electrons and positrons~\cite{Rohrlich1954} could contribute to the difference we see between pair production and electrons.


\subsubsection{Energy resolution achieved}


Detailed understanding of $\beta$ energy spectra requires knowledge of the energy resolution, which must be folded into the simulation physics. Typically the energy resolution of plastic scintillator with PMT readout can be assumed to be dominated by photon statistics, requiring one constant to model accurately~\cite{Fenker2018}. SiPM readout should introduce a contribution from dark current that would be independent of $\beta$ energy.
We determine the dark current contribution by measuring the plastic scintillator detector noise when the detector has no event above threshold, i.e. from the QDC pedestal. The result is constant energy noise of 31 keV at one $\sigma$, constant with energy.

\begin{figure}[htb]
   \includegraphics[angle=90,width=\linewidth]{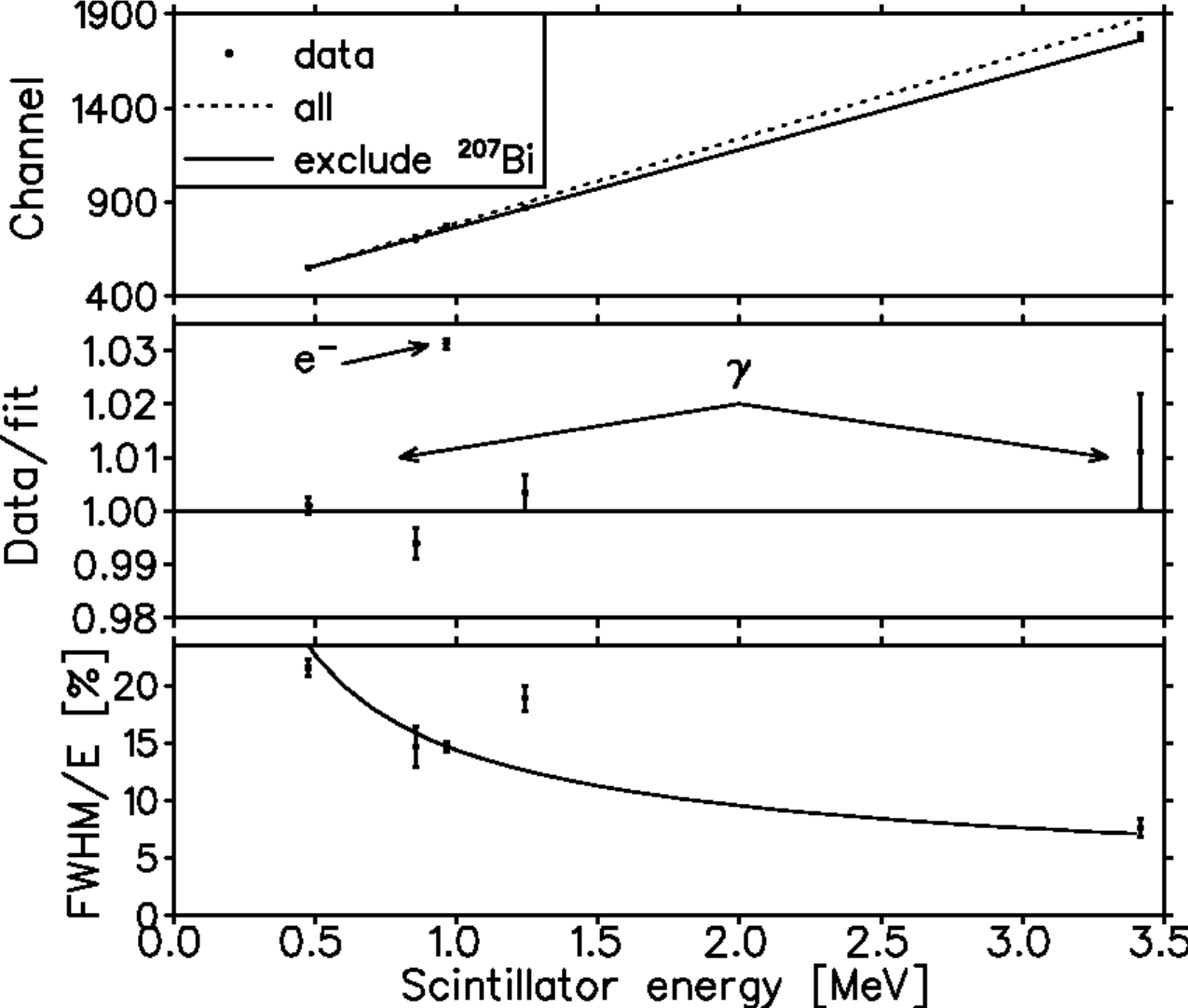}

  \caption{
    Top:    Deviations from linearity of the energy calibration. The $^{207}$Bi IC electron point excluded is the one at 995 keV. 
    Middle: Residuals of top plot, i.e. data divided by the fit with $^{207}$Bi excluded.
        Bottom: Energy resolution fit to photon statistics added in quadrature with dark current.
          }
  \label{fig-linearity}
\end{figure}

The energy resolution of the plastic scintillator system as a function
of energy is shown in Fig.~\ref{fig-linearity}, using the restricted Gaussian fit above for the 3.42 MeV point.
We determine the energy resolution for Compton edges by convolution of a Gaussian with a GEANT4 simulation. 
We show a fit assuming photon
statistics added in quadrature with dark current, with result
FWHM/E[MeV] = $\sqrt{(12.5\pm1.1\%/\sqrt{\rm E[MeV]})^2 + (7\%/E[MeV])^2}$.
The result would not be changed significantly by arbitrarily excluding the discrepant $^{40}$K Compton edge.
The single high-energy point is clearly very helpful to determine the energy dependence of the resolution.
Whether it's better to determine the energy resolution from measurements like these, or whether it's better to fit one or two parameters as part of a fit of physics to the $\beta$ spectrum~\cite{Fenker2018}, depends on details of the experiment.

\subsubsection{Lineshape response and GEANT4 simulation}

A GEANT4 simulation for 4.44 MeV $\gamma$'s is also shown in Fig.~\ref{fig-plastic}, using the emstandard$_{opt3}$ physics list as in Ref.\cite{Fenker2018}.
The simulation shows a tail on the response function, a 
combination of bremsstrahlung escape from the energetic e$^+$e$^-$ pair and
pairs created near the edge of the scintillator. The simulated
tail below 90\% of the peak has tail/peak ratio 0.13, with relatively fewer events at lower energy. The experimental tail/peak is 0.94$\pm$0.10, an order-of-magnitude larger, with relatively more events at lower energy.



Also shown in Fig.~\ref{fig-plastic} (Bottom) is the background energy
spectrum from events with higher or lower energy in one GAGG scintillator,
which also shows a substantially larger number of events at lower scintillator energy.
So we can
attribute most of the low-energy tail to neutron-produced events mimicking the
511 coincidence, but depositing less energy than the 3.42 MeV pair production peak. The large size of the correction needed for a valid comparison, along with the poor statistics, makes a meaningful correction impossible with the present data set. Although in principle modern simulations can include neutron-induced events, we do not consider the much greater complication here. 

\section{Summary and Possible Improvements}
In summary, by triggering on annihilation radiation, we have measured
the centroid and FWHM resolution of a 40x88 mm  plastic scintillator at 3.42 MeV,
the double-escape peak from 4.44 MeV $\gamma$'s from an AmBe neutron source. We have quantified systematic uncertainty from a non-Gaussian tail on the double escape peak.

The precision of the resulting centroid shows a systematic difference between $\gamma$ (i.e. average of whole detector) and electron (centered on
front face) calibrations. This is consistent with our simulations of nonuniform light collection, though the known difference in e$^-$ and e$^+$ energy loss in scintillators~\cite{Rohrlich1954} could also contribute. Thus am empirical correction, in our case 3\% at 1 MeV,
would have to be made to a $\gamma$-based calibration to use this detector
for $\beta$ spectroscopy.

The FWHM at 3.42 MeV, combined with calibrations at lower energies, is modelled reasonably consistently with photon statistics and a measured contribution from dark current. Depending on accuracy needed, most experiments would want to constrain parameters to fit to $\beta$ spectra by making such dedicated measurements.




Our main goal was the measure the lineshape tail, the most important feature to fold into theory to
use the detector for $\beta$ spectroscopy, because it is the largest effect misidentifying high-energy $\beta$ strength. The neutron backgrounds are at least an order of magnitude too large to make a valid test of the GEANT4 simulation, limited by several aspects of our implementation of this technique. So we consider possible improvements. Scintillation light readout insensitive to neutron damage, e.g. PMT's instead of SiPM's, could allow higher-statistics data collection. Order-of-magnitude better timing is also possible with either PMT or more segmented SiPM readout and faster high-Z scintillator, which could cut some fraction of fast neutron-produced events. The plastic scintillator could be replaced by similarly low-Z stilbene or organic glass scintillator to reject neutron events with pulse-shape discrimination~\cite{ogs}.


{\bf Acknowledgements} We acknowledge B.L. Lee for a previous generation of this experimental technique, and J.C. McNeil and B. Kootte for constructive comments. Supported by RBC Foundation and the Natural Sciences and Engineering Research Council of Canada. TRIUMF receives federal funding via a contribution agreement through the National Research Council of Canada.


\begin{thebibliography}{00}

  \bibitem{Fenker2018}
 ``Precision Measurement of the $\beta$ Asymmetry in Spin-Polarized $^{37}$K Decay,''
  B. Fenker, A. Gorelov, D. Melconian, J.A. Behr, M. Anholm, D. Ashery, R.S. Behling, I. Cohen, I. Craiciu, G. Gwinner, J. McNeil, M. Mehlman, K. Olchanski, P.D. Shidling, S. Smale, and C.L. Warner,
  Phys. Rev. Lett. {\bfseries 120}, 062502 (2018)

\bibitem{Minamisono2011} ``Low-energy test of second-class current in $\beta$ decays of spin-aligned $^{20}$F and $^{20}$Na,'' K. Minamisono, T. Nagatomo, K. Matsuta, C.D.P. Levy, Y. Tagishi, M. Ogura, M. Yamaguchi, H. Ota, J.A. Behr, K.P. Jackson, A. Ozawa, M. Fukuda, T. Sumikama, H. Fujiwara, T. Iwakoshi, R. Matsumiya, M. Mihara, A. Chiba, Y. Hashizume, T. Yasuno, and T. Minamisono,
  Phys Rev C {\bfseries 84}, 055501 (2011).

\bibitem{Clifford1983} ``Kinematic Shifts in the $\beta$-Delayed Particle Decay of $^{20}$Na, and the $\beta$-$\nu$ Angular Correlation,'' E.T.H. Clifford, J.C. Hardy, H. Schmeing, R.E. Azuma, H.C. Evans, T. Faestermann, E. Hagberg, K.P. Jackson, V.T. Koslowsky, U.J. Schrewe, K.S. Sharma, and I.S. Towner,
  Phys. Rev. Lett. {\bfseries 50} 23 (1983)

\bibitem{Clifford1984} ``Measurements of the Response of a Hybrid Detector Telescope to Monoenergetic Beams of Positrons and Electrons in the Energy Range 0.8-3.8 MeV,'' E.T.H. Clifford, E. Hagberg, V.T. Koslowsky, J.C. Hardy, H. Schmeing, and R.E. Azuma,
  Nucl. Inst. and Meth. {\bfseries 224} 440 (1984)

\bibitem{Lorch1973}
  ``Neutron spectra of $^{241}$Am/B, $^{241}$Am/Be, $^{241}$Am/F, $^{242}$Cm/Be, $^{238}$Pu/$^{13}$C, and $^{252}$Cf isotopic neutron sources,''
  E.A. Lorch, Int. J. Appl. Radiat. Isot. {\bf 24} 58 (1973)

\bibitem{Scherzinger2017}
``A comparison of untagged gamma-ray and tagged-neutron yields from $^{241}$AmBe and $^{238}$PuBe sources,''
    J. Scherzinger, R. Al Jebali, J.R.M. Annand, K.G. Fissum, R.Hall-Wilton, S.Koufigar, N. Mauritzson, F. Messi, H. Perrey, and E. Rofors,
    Applied Radiation and Isotopes, {\bf 127} 98 (2017).
    
  \bibitem{Ghosh2020}
``Dependence of polytetrafluoroethylene reflectance on thickness at visible and ultraviolet wavelengths in air,''
    S. Ghosh, J. Haefner, J. Martín-Albo, R. Guenette, X. Li, A.A. Loya Villalpando, C. Burch, C. Adams, V. Álvarez, and L. Arazi,
    2020 JINST 15 P11031

  \bibitem{Haefner2017}
    ``Reflectance dependence of polytetrafluoroethylene on thickness for xenon scintillation light,''
J.Haefner, A.Neff, M.Arthurs, E.Batista, D.Morton, M.Okunawo, K.Pushkin, A.Sander, S.Stephenson, Y.Wang, W.Lorenzon,
   Nucl. Inst. and Method. A {\bf 856} 86 (2017)
       

  
\bibitem{nitrocellulose}
 ``Reflectivity Spectra for Commonly Used Reflectors,''
  M. Janecek,
  IEEE Transactions on Nuclear Science, {\bf 59} 490 (2012)
  

\bibitem{Musienko2017}
 ``Radiation damage in silicon photomultipliers
exposed to neutron radiation,''
  Yu. Musienko,
  A. Heering, R. Ruchti, M. Wayne, Yu. Andreev, A. Karneyeu, and
  V. Postoev, 2017 JINST 12 C07030

  \bibitem{Tsang2016}
``Neutron radiation damage and recovery studies of SiPMs,''
T. Tsang, T. Rao, S. Stoll and C. Woody,
2016 JINST 11 P12002

\bibitem{differential}
``Differentiation method for localization of Compton
edge in organic scintillation detectors,''
M.J. Safari, F. Abbasi Davani, H. Afarideh, arXiv:1610.09185.

\bibitem{Rohrlich1954}
  ``Positron-electron differences in energy loss and multiple scattering,''
  F. Rohrlich and B.C. Carlson,
  Phys. Rev. {\bf 93} 38 (1954).

\bibitem{FenkerThesis} ``Precise Measurement of the $\beta$-asymmetry in the
  Decay of Magneto-optically Trapped, Spin-Polarized $^{37}$K,''
  Benjamin Brown Fenker, Ph.D. thesis, Texas A\&M, 2016.

\bibitem{ogs}
  ``Organic glass scintillator (OGS) property comparisons to Stilbene, EJ-276 and BC-404,''
  W.K. Warburton, J.S. Carlson, and P.L. Feng,
  Nucl. Inst. and Meth. A 1018 165778 (2021)
  

  \end{thebibliography}
\end{document}